# Ferromagnetism in graphene nanoribbons: split versus oxidative unzipped ribbons

S. S. Rao[§,*], S. Narayana Jammalamadaka[&,*], A. Stesmans[§] and V. V. Moshchalkov[&]

[§]*INPAC – Institute for Nanoscale Physics and Chemistry, Semiconductor Physics Laboratory, K.U. Leuven, Celestijnenlaan 200D, B–3001 Leuven, Belgium*

[&]*INPAC – Institute for Nanoscale Physics and Chemistry, K.U. Leuven, Celestijnenlaan 200D, B–3001 Leuven, Belgium*

J. van Tol[a]

[a]*National High Magnetic Field Laboratory, Centre for Interdisciplinary Magnetic Resonance, Florida State University,1800 E. Paul Dirac Drive, Tallahassee, Florida 32310, USA*

D. V. Kosynkin[1], A. Higginbotham[1] and J. M. Tour[1,2,3]

[1]*Department of Chemistry, [2]Department of Mechanical Engineering and Materials Science, [3]Smalley Institute for Nanoscale Science and Technology, Rice University, MS-222, 6100 Main Street, Houston, Texas 77005, USA.*

***\* These authors contributed equally to this work***

Two types of graphene nanoribbons: (a) potassium-split graphene nanoribbons (GNRs), and (b) oxidative unzipped and chemically converted graphene nanoribbons (CCGNRs) were investigated for their magnetic properties using the combination of static magnetization and electron spin resonance measurements. The two types of ribbons possess remarkably different magnetic properties. While the low temperature ferromagnet-like feature is observed in both types of ribbons, such room temperature feature persists only in potassium-split ribbons. The GNRs show negative exchange bias, but the CCGNRs exhibit a 'positive exchange bias'. Electron spin resonance measurements infer that the carbon related defects may responsible for the observed magnetic behaviour in both types of ribbons. Furthermore, proton hyperfine coupling strength has been obtained from hyperfine sublevel correlation experiments performed on the GNRs. Electron spin resonance provides no indications for the presence of potassium (cluster) related signals, emphasizing the intrinsic magnetic nature of the ribbons. Our combined experimental results may infer the coexistence of ferromagnetic clusters with anti-ferromagnetic regions leading to disordered magnetic phase. We discuss the origin of the observed contrast in the magnetic behaviours of these two types of ribbons.

*Corresponding authors

srinivasasingamaneni@boisestate.edu , Surya.Jammalamadaka@fys.kuleuven.be

Graphene is a promising material for spintronic applications due to its high carrier mobility at room temperature and its long spin life time resulting from low spin-orbit coupling ($\sim 10^{-4}$ eV) and low hyperfine interaction [1,2]. Graphene is a zero band gap semiconductor, and it has been recognized that the graphene nanoribbons (GNRs) may be the potential candidates in future MOSFET devices with excellent ON/OFF ratios over graphene as the band gap of the GNRs can be tuned as a function of ribbon width, quantum confinement, edge geometry, doping and edge functionalization. GNRs present long and reactive edges prone to localization of electronic edge states and covalent attachment of chemical groups can significantly alter their electronic and magnetic properties [3].

It has been predicted that zigzag-edged graphene or semi-hydrogenated graphene sheets would exhibit ferromagnetic and spintronic properties [1]. Peculiar magnetic properties such as flat band ferromagnetism (FM) in zigzag GNRs, and the presence of spin waves and dynamical edge state magnetism have also been theoretically predicted [4,5]. Recent theoretical studies have indicated that the magnetic properties of zigzag GNRs can be controlled by an external electric field and predicted that edge states are strongly spin polarized [6]. In addition to that, the occurrence of carrier induced FM in GNRs has also been envisaged [7]. Theoretical work suggests the presence of strong ferromagnetic interaction ranging several $10^3$ K between the edge-state spins in the zigzag edge that can create carbon only FM [8]. The most recent and outstanding works on these materials showed that the magnetic properties are not exclusively related to the presence of extrinsic magnetic ions but strongly determined by the defects. For example, in pristine graphite, proton irradiation has been found to be an effective way to produce bulk FM [9]. While bulk graphene is known to be non-magnetic, its derivatives such as graphene oxides, few-layer hydrogenated epitaxial graphene, and hydrogenated carbon nanotubes (CNTs) are reported to show room temperature

FM (RTFM) [10-12]. Interestingly, a recent scanning tunneling microscopy/spectroscopy of GNRs with ultra-smooth edge states with characteristic splitting in the dI/dV spectra has been reported – an unambiguous indication of magnetic ordering [13].

However, the direct assignment and attribution of the observed RTFM has been hampered in bulk static magnetization measurements as they provide cumulative magnetic response. One of the main supporting arguments for the observation of intrinsic FM in carbon-derived materials was obtained from X - ray magnetic circular dichroism (XMCD) and magnetic force microscopy (MFM) as intrinsic defects [14]. In addition to that, the importance of single atom defects such as carbon (defect) contribution and the role of adatoms such as hydrogen to the observed FM are still being investigated and widely discussed [1,11,12,15]. Despite several interesting magnetic properties of GNRs have been predicted theoretically as outlined above, their experimental realization has been at the nascent stage [16,17]. Therefore, to understand the graphene-based magnetism more thoroughly, further sensitive experimental verification is required. Particularly, to use GNRs in spin-based applications, it is an utmost important to investigate their true magnetic nature and the species (spins) that cause the magnetic order.

For the present work, we have chosen two sets of GNRs, characterized by long and straight edges. With the advent of recent chemical approaches [18,19], gram-scale GNRs could be produced, thereby facilitating the investigation of their magnetic properties. To probe the magnetic behaviour of the ribbons, we use classical static magnetization measurements coupled with defect-type-specific electron spin resonance (ESR) technique and hyperfine sublevel correlation (HYSCORE) spectroscopy. In this paper, we report on important findings pertaining to the magnetic nature of two types of GNRs, namely those derived by potassium splitting of multi walled carbon nanotubes (MWCNTs) to produce GNRs and oxidative unzipped MWCNTs that produce graphene oxide nanoribbons which were further reduced by hydrazine, named as chemically converted graphene nanoribbons (CCGNRs). Though both the ribbons exhibit low temperature ferromagnetlike (FM-like) behaviour, such room temperature feature persists only in the GNRs. The isothermal magnetization behaviour could be well described by the Langevin function in both the ribbons, typically applicable for disorder magnetic materials such as superparamagnets. The GNRs show negative exchange bias while the CCGNRs ribbons show positive exchange bias, the later behaviour is unusual. From conventional ESR and HYSCORE data, we infer that carbon-defect related clusters may cause the observed magnetic behaviour in the GNRs. The present results deserve special attention as the GNRs exhibit FM-like behaviour without applying external perturbations such as doping, hydrogenation, applying external electric field, ion implantation or high energy bombardment. Particularly, the present results are linked to technological interest in spintronic based applications with the RTFM-like properties coupled with their semiconducting properties.

The synthesis of K-split GNRs (GNRs) involves [18] heating potassium metal and MWCNTs with outside diameter of 40-80 nm and approximately 15-20 inner nanotube layers are sealed in a glass tube and heated in a furnace at 250 $^o$C for 14 h, followed by quenching to effect the longitudinal splitting process. The splitting process might be further assisted by the generation of $H_2$ upon the ethanolic quench. The split MWCNTs are further exfoliated to form GNRs upon sonication in chlorosulfonic acid. This procedure produces GNRs free from oxidative damage with conductivities paralleling the properties of the best samples of mechanically exfoliated graphene. Scanning electron microscopy (SEM) and atomic force microscopy (AFM) show that the GNRs have widths of 130-250 nm and a length of 1-5 μm. Several techniques were used to fully characterize GNRs and to test their electronic properties, as reported elsewhere [18]. CCGNRs were prepared by longitudinal unzipping of MWCNTs [19]. Briefly, this method involves the treatment of MWCNTs, consisting of 15-20 concentric cylinders and 40-80 nm diameter with concentrated $H_2SO_4$ followed by $KMnO_4$ (an oxidizing agent) at room temperature, consequently reduced by $N_2H_4$, named as chemically converted graphene nanoribbons (CCGNRs). From very preliminary studies, both ribbon types have mainly zigzag edges with some short domains of armchair edges.

From a broader context, the ribbons made from potassium are far more conductive as their planes remain pristine throughout the chemistry. For example, these ribbons, as we have published [18], are as conductive as exfoliated graphene from graphite. They are almost never monolayered as they remain foliated in stacks of about 5-8 thick since the planes are pristine. The edge anions from the reaction were quenched with a proton source, hence hydrogen atoms likely edge the ribbons—though precise confirmation of such a claim is almost impossible using any analytical method available today.

The ribbons made from oxidative unzipping produce graphene oxide nanoribbons (GONRs) that are heavily oxidized both on the planes and at the edges due to the strongly oxidative conditions. Hence, they are poorly conductive but very well exfoliated in the aqueous system in which they are prepared, thereby often being monolayered. For the studies here they were subsequently hydrazine reduced. Hydrazine will remove about 95% of the oxygen functionalities [19]. This restores much of the conductivity, but they are still higher in resistivity than the potassium-derived ribbons noted above. Again, the I-V curves of the two types are well covered in the former papers [18,19]. Since these are now chemically converted graphene (CCG) nanoribbons, they have holes and some remaining oxygen functionalities, predominantly on the edges. The increased and remaining edge structures in the oxidative unzipped structures can indeed affect the measurements here.

Static magnetization measurements were performed using vibrating sample magnetometer (VSM) equipped with Oxford cryostat on tightly packed GNRs and CCGNRs in non-magnetic straw. Conventional first derivative CW ESR experiments were carried out in the temperature (T) range of 4.2 – 120 K using a home built K-band ($\approx$ 20.6 $GHz$) ESR spectrometer operated under conditions of adiabatic slow passage. Conventional low power first-derivative-absorption d$P_{\mu r}$/dB ($P_{\mu r}$ being the reflected microwave power) spectra were detected through applying sinusoidal modulation ($\sim$ 100 kHz, amplitude $B_m \sim 0.42 \times 10^{-4}$ T) of the externally applied magnetic field $\vec{B}$, with incident microwave power $P_\mu$ as well as $B_m$ cautiously reduced to avoid signal distortion. The defect spin density was quantified by double numerical integration of the $K$-band derivative absorption spectra by making use of a co-mounted calibrated Si:P intensity marker, also serving as g marker: $g$ (4.2 K = 1.99869) [20]. X - band ($\sim$ 9.7 $GHz$) HYSCORE spectroscopy was carried out at T = 12 K with a repetition time of 1 $ms$ using the sequence $\pi/2$-$\tau$-$\pi/2$-$t_1$-$\pi$-$t_2$-$\pi/2$-$\tau$-echo. The mw pulse lengths were $t_{\pi/2}$ = 16 $ns$ and $t_\pi$ = 28 $ns$, starting times $t_{10}$ = $t_{20}$ = 260 $ns$, a $\tau$ = 164 and a time increment of $\Delta$t = 8 $ns$ (data matrix 256 × 256). Special attention has been paid to make sure that no external impurity was introduced while sample handling during the measurements.

We begin our discussion with the $dc$ magnetization measurements which are summarized in Figs.1-3. Fig. 1a displays the isothermal magnetization data measured on GNRs at 5, 10, 50, 100, 200 and 300 K with the magnetic field varied from -5 to + 5 T. A clear, s-shape saturated open hysteresis loops (M (H) curve) were observed at all the temperatures measured including at 300 K, with the coerceivity ($H_C$) of $\sim$ 0.02 T and the saturated magnetization of ($M_S$) $\sim$ 0.25 emu/g. These numbers are comparable to those reported for some of the carbon based materials [9-12]. From the main panel of Fig.1(a), it can be noted that the saturation magnetization $M_0$ decreases as the temperature increases from 5 to 300 K, a typical signature of nominal FM-like materials. The top left inset of Fig. 1(a) describes the enlarged view of M-H curves measured on GNRs. The isothermal magnetization data at 5 K could be satisfactorily fitted with a Langevin function $L(x)$ valid for classical spins of $S = \frac{1}{2}$, typically describes the disordered magnetic behaviour such as superparamagnetism, given as

$$M = M_0 \left[ \coth\left( \frac{\mu_0 \mu_B H}{k_B T} \right) - \left( \frac{\mu_0 \mu_B H}{k_B T} \right)^{-1} \right] \qquad (1)$$

Where $M$ is the magnetization, $M_o$ is the saturation magnetization in emu/g, $\mu_B = 9.274 \cdot 10^{-24}$ J/T is the Bohr magneton, $k_B = 1.381 \cdot 10^{-23}$ J/K is the Boltzmann constant, T = 5 K is the temperature and $H$ is the magnetic field in Tesla. As shown in the lower right inset of Fig.1.(a), a fit to the experimental data leads to a magnetic moment of 34.27 (5) $\mu_B$ suggesting a clustering of the $S = \frac{1}{2}$ spins. The FM-like phase is found to be highly stable over a period of one year as inferred from our isothermal M-H data (not shown). Such $sp$-electron magnetic clustering has been predicted [21] theoretically from graphene-based materials, also inferred from other carbon-based materials such as carbon nanofoam [22]. In a recent theoretical work [23], such carbon adatom cluster/agglomeration induced FM-like phase has been demonstrated in graphene. Fig.1.(b) shows isothermal M (H) plots recorded from CCGNRs measured at various temperatures 1.8, 5, 10 and 15 K. The typical signature of FM-like phase can be inferred from CCGNRs at 1.8 K. However, as the temperature increases the FM-like phase seems to disappear. This particular observation can be seen more clearly from the zoomed spectrum, as displayed in the top left inset of Fig. 1(b). These ribbons show prominent FM-like features (open saturated MH hysteresis loop with well-defined coerceivity) at 1.8 K and gradually the paramagnetic feature (more linear MH) takes over as the temperature rises to 10 K. Similar to the case of GNRs, the isothermal magnetization data of CCGNRs at 1.8 K could be satisfactorily fitted with a Langevin function $L(x)$ (c.f. equation (1)) valid for classical spins of $S = 1/2$ as shown in the lower right inset of Fig.1(b), leading to a magnetic moment of 7. 04 (6) $\mu_B$ suggesting a clustering of S = $\frac{1}{2}$ spins. As one can notice, the effective magnetic moment in the CCGNRs is almost 4.8 times smaller than GNRs which might explain the apparent room temperature FM-like feature of GNRs. Such disordered magnetic feature from graphene based materials has been discussed widely in the literature [2,9-12,24]. The apparent absence of such room temperature FM-like feature in CCGNRs can be explained as the majority of the edges are reconstructed, passivated or closed [25] upon the oxidative unzipping process. Only a few remaining magnetic edges would give rise to the weak paramagnetic contribution at room temperature [25].

The results of temperature-dependent magnetization measurements for GNRs and CCGNRs recorded at 0.1 T are shown in Figs.2 (a,b). First measurement was taken after zero-field cooling (ZFC) to lowest temperature possible (4.4 K) and in the second run the measurements were taken under field-cooled (FC) conditions. From these plots, several salient features can be inferred : low temperature susceptibility (cf. Fig.2a) of the former is about 3 times higher than the latter (cf. Fig.2b).GNRs show pronounced FC dependence: $\chi$ (T) is larger under FC conditions when compared to values measured under ZFC conditions (cf. Fig. 2a). While FC dependence typically shows monotonic increase of magnetic susceptibility with decreasing temperature, the ZFC data (cf. Fig.2a) displays much more complex behaviour. The ZFC data exhibits a broad maximum at around ~ 70 K and low temperature sharp upturn at around ~ 14 K upon progressive cooling. The existence of maxima in ZFC data can be a consequence of competing interactions and/or low dimensionality of the magnetic system. The presence of small magnetic domains or small regions with a super magnetic behaviour may also give rise to such a peak. However, the ZFC stationary points are absent in CCGNRs (cf. Fig.2b). Moreover, while cooling from room temperature, both ZFC and FC data follow similar trend, i.e., slow increase of susceptibility untill ~ 30 K followed by a sharp rise. We note here a striking similarity between the ZFC susceptibility measured in GNRs with that of modified graphite and carbon nanofoam [26,27].

To probe the magnetic behaviour of the ribbons further, isothermal M-H measurements were performed on GNRs and CCGNRs at 5 K under ZFC and FC conditions. M (H) loops are plotted in Fig. 3a for GNRs measured under ZFC and FC (FC: 0.1, 0.3, 0.5 and 1 T) conditions. Upon closer inspection, as displayed in the top left inset for clarity, upon field cooling, M (H) loops shift towards negative magnetic field axis – a phenomenon known as 'negative exchange bias' (NEB) – an usual phenomenon that has been commonly observed in systems exhibiting co-existing magnetic phases such as ferromagnetism (FM) and anti-ferromagnetism (AFM). Interestingly, in the case of CCGNRs, as shown in Fig.3b, the FC (FC: 0.1, 0.2, 0.4, 0.6 T) causes M (H) curves to shift towards positive field axis -so called 'positive exchange bias' (PEB), as clearly displayed in the top left inset of the Fig.3b. Such PEB is unusual for metal free systems such as CCGNRs though it has been reported earlier in bilayer metallic systems such as NiFe/IrMn and $FeF_2$-Fe bilayers [28,29], and in $La_{0.67}Sr_{0.33}MnO_3/SrRuO_3$ [30].

This particular part of our work probes the interactions develop at the magnetic interfaces in two types of graphene nanoribbon samples. Probing such interactions is important to understand the intrinsic magnetic behaviour of the ribbons. Analogous to the case of intrinsic interface exchange coupling between FM and AFM phases in phase separated charge ordered manganites [31], we studied the cooling field ($H_{FC}$) dependence of exchange bias ($H_{EB}$) on GNRs and CCGNRs. Now we define the exchange bias field $H_{EB}$ = -($H_{right}$+$H_{left}$)/2, $H_{right}$ and $H_{left}$ being the points where the loop intersects the field axis. Let us analyze the field-cooling dependence of the loop parameters $M_r$, $H_{EB}$ and $H_C$ at T = 5 K for both GNRs and CCGNRs.

We did not find significant vertical loop shift (variation of $M_r$) as a function of field cooling (data not shown) for both ribbons. The variation of $H_{EB}$ obtained at 5 K as a function of cooling field for both GNRs and CCGNRs is plotted in the right bottom insets of the Figs.

(3a,b), respectively, which deserves special attention. In both GNRs and CCGNRs, the $H_{EB}$ increases strongly and peaks at ~ 0.4 T and ~ 0.2 T, respectively, and then falls as a function of further increase in cooling field. Similar observations were reported earlier (cf. Fig.4b of [32] ) for a granualar system of Fe nanoparticles embedded in an Fe oxide matrix. The authors in [32] have explained their observation by invoking the complex interplay of AFM-FM (super)-exchange coupling and particle-particle dipolar interactions, which both can be FM or AFM in sign. We shall explain this interesting observation more quantitatively in forthcoming paragraph.

Figure 4 describes the variation of coercivity ($H_C$) as a function of $H_{FC}$ for (a) GNRs (b) CCGNRs measured at 5 K. As it can be noticed, the increase in $H_c$ with field cooling up to ~ 0.4 T (in GNRs) and ~ 0.2 T (in CCGNRs) is coherent with the increase in $H_{EB}$ (cf. Right bottom insets of Figs.3(a,b)): the field-cooling induces a preferential direction along which the magnetic moments tend to freeze at 5 K and thus the effect of averaging of anisotropy is reduced.  As the cooling field increases up to ~ 0.4 T (in GNRs) and ~ 0.2 T (in CCGNRs) , the exchange anisotropy at the interface between the FM-AFM regions sets in when the ribbons are cooled down to 5 K and hence a large increase in $H_{EB}$ is observed. At higher cooling fields, the effective magnetic coupling (Zeeman coupling) between the field and the AFM moments increases which further orients AFM moments along the field direction. At that stage, the Zeeman coupling overcomes the exchange coupling at the interface of FM-AFM, thus leading to decrease in $H_{EB}$. A similar explanation has been proposed to account for the positive exchange bias observed in $FeF_2$ (AFM)-Fe(FM) bilayers [29].

Analogous to phase separated perovskite manganites, the variation of $H_{EB}$ as a function of cooling field ($H_{FC}$) can be expressed by the following relation [31]

$$-H_{EB} \propto J_i \left[ \frac{J_i \mu_0}{(g \mu_B)^2} L \left( \frac{\mu H_{FC}}{k_B T_f} \right) + H_{FC} \right] \qquad (2)$$

$J_i$ is the interface exchange coupling constant, $g$ is the gyromagnetic factor, $L(x)$ is the Langevin function, $x = \mu H_{FC}/k_B T_f$, $\mu_B = 9.274 \cdot 10^{-24}$ J/T is the Bohr magneton, $k_B = 1.381 \cdot 10^{-23}$ J/K is the Boltzmann constant and $\mu$ is the magnetic moment of the clusters, $H_{FC}$ is cooling field and $T_f = 5$ K. From the bottom right insets of Fig. 3(a,b) it can be seen that the experimental data could be well described by above model as shown by solid curve. The obtained values for the $J_i$ for both the GNRs and CCGNRs are ~ - 95 *meV* and -8.4 *meV*, respectively. We also estimated the Curie temperatures $T_C$ using the mean-field expression $T_C = JS(S+1)/3k_B$ [33] for $S = \frac{1}{2}$, and obtained as $T_C = 276$ K and 24 K, for GNRs and CCGNRs, respectively. These numbers are in close agreement with our experimental observations of ~ 300 K and 10 K for GNRs and CCGNRs, respectively.

According to the literature reports [28-30], for PEB systems, i.e) similar to CCGNRs, the sign of $J_i$ is found to be negative inferring the anti-ferromagnetic exchange coupling exists between FM and AFM phases. However, in the case of systems similar to GNRs, which show NEB, it was reported that the sign of $J_i$ can be a positive or negative. The observed –ve $J_i$ for GNRs implies the presence of anti-ferromagnetic exchange coupling between co-existing FM and AFM phases. The observation of PEB in CCGNRs could be attributed to their rough micro-structure of the ribbons resulted from heavy oxidative damage due to harsh acidic treatment during the unzipping process, as directly evidenced from the transmission electron microscopic image [19]. Such rough structure is absent in GNRs [18] which may result in smooth edges facilitating NEB.

Fig. 5 shows representative low-power (2.5 *nW*) *K*-band ESR spectra measured on GNRs (a) and CCGNRs (b) at 4.2 K. As shown, only one ESR signal is observed at zero-crossing $g$ value ($g_c$) = 2.0025 and 2.0032 , respectively. These numbers are close to free electron $g$ value, 2.0023 inferring that the observed signal does not originate from transition metals. This value falls within the reported [34] carbon ESR signal range ($g$ = 2.0022 - 2.0035), indicating the signal may be ascribed to $C$ related dangling bonds of spin $S = \frac{1}{2}$. For the GNRs, the signal deviates from a Lorentzian of peak-to-peak width of $\Delta B_{PP} \sim 8 \times 10\text{-}4\ T$ with the spin density of $\sim 6 \times 10^{19} g^{-1}$. For the CCGNRs, the signal closely follows the Lorentzian, having $\Delta B_{PP} \sim 6 \times 10\text{-}4\ T$ with the spin density of $\sim 1.5 \times 10^{17} g^{-1}$. One can notice that the number of spins in the former case is more than two orders of magnitude higher than in the latter. Despite intense signal averaging over broad field ranges under various extreme and optimized spectrometer parameter settings, no other signals could be observed over a broad magnetic field sweep range up to $0.9T$. Though intensely searched for, neither any correlated additional signal structure could be traced nor any sign of hyperfine structure possible ensuing from $^1$H, $^{14}$N, $^{19}$K, $^{16}$S, or $^{55}$Mn nuclei. Aforementioned data and discussed experimental results refute suggestions that ferromagnetism in GNRs arises only from the precipitation of metals. Instead, the data presented in Figs 1 - 5 offer clear evidence for the importance of factors other than the presence of transition-metals in governing the magnetism of these ribbons.

Information on possible magnetic nuclei (if any) coupled to the paramagnetic centers can be obtained with hyperfine sublevel correlation (HYSCORE) spectroscopy carried out at 12 K and at 9.7 GHz on the GNRs. HYSCORE is a high-resolution 2-dimensional pulsed ESR technique used here to measure small hyperfine (hf) couplings not resolved in the CW ESR spectrum [35]. The HYSCORE spectrum shown in Fig.6 allows to identify two nuclear isotopes coupled to the electron spins. There is a strong peak at 14 *MHz* which is the NMR frequency of protons, which corresponds to weakly coupled protons, and a weaker peak at around 3.5 *MHz*, which corresponds to the NMR frequency of $^{13}C$. Figure 6 displays ridges along the proton diagonal line with proton hyperfine couplings in the range of 20-28 *MHz* (~0.890 mT), implying that some radicals are coupled to protons. The nonzero isotropic component of the hyperfine coupling results from a Fermi contact interaction, which indicates an electronic connection between the unpaired electron and the proton(s), rather than a pure through space dipole-dipole interaction. A single carbon centered radical with bonded hydrogen atoms like a methyl radical can be ruled out since in this case typically A($^1$H) ~ 67 *MHz* (2.4 mT), a hyperfine coupling much bigger than observed in the HYSCORE spectrum and would result in an observable splitting in the GNRs CW ESR spectra (line width 0.6 mT), contrary to the present observation. HYSCORE data from Fig.6 reveals proton hyperfine couplings around 25 *MHz*, a value comparable to those of conjugated aliphatic radicals [36]. It is likely that these hyperfine couplings are related to protons on the edges of the nanoribbons, in line with the characterization data [18] which show the termination of edges with protons as a result of quenching of aryl potassium edges with ethanol.

Similar to the analysis reported [37] on hydrogen absorption in ball milled graphite, if we assume the ESR signals result from $\pi$ radicals, then we can use the proton hyperfine couplings to estimate the electron spin density $\rho^\pi$ at the adjacent carbon atoms with the McConnell equation [38], $a_H(\alpha) = Q_H\rho^\pi$, where $Q_H$ ~ 2.7 mT, is an intrinsic coupling for unit density and $a_H(\alpha)$ is the proton hyperfine coupling in mT. The hyperfine coupling of ~ 0.89 mT from the HYSCORE, thus corresponds to a carbon electron spin density about 33%. Interestingly, this is only slightly higher as the predicted [39-41] spin densities of localized edge states in zigzag graphene nanoribbons.

In addition to the proton hyperfine interactions, the HYSCORE $^{13}C$ signals (in natural abundance) at low frequencies in Fig.6 indicate $^{13}C$ hyperfine couplings in the range |A($^{13}C$)| < 4 *MHz* (0.14 mT). These small couplings are clearly not in line with the above mentioned spin density of 0.33, which would result in much larger $^{13}$C couplings [42]. However, these large couplings with large anisotropy coupled with the low [13]C natural abundance will be difficult to detect in a HYSCORE experiment on a non-oriented sample. The observed [13]C hyperfine splitting can be due to [13]C nuclei closer to the center of the nanoribbons with significantly lower spin density [40,42]. From the results of HYSCORE experiments, it is possible that the larger width ($\sim 8 \times 10^{-4}$T) of the ESR signal from GNRs in comparison to that ($\sim 6 \times 10^{-4}$T) of the CCGNRs can be due to the protons present in the GNRs.

Hence the hyperfine couplings derived from the HYSCORE data show that the (unpaired) spin density, and thus wave function of the radical is spread over small graphitic moieties. Most likely the magnetic states are created from the cleavage of graphene sheets of MWCNTs, during the splitting process, resulting in so-called zigzag and armchair edges. There could also be a population of radicals associated with aliphatic structures. These structural types are thought to create nonbonding $\pi$-electron states, which accommodate a high unpaired spin concentration. The large spatial but still localized wave functions of the radicals allow exchange interactions of varying strengths sufficient to induce ferromagnetic order. From our experimental results, we suggest that the clustering of magnetic fragments comprises of carbon defects coupled with hydrogen could be one of the possible sources of the ferromagnetism. From our study, it is clearly evident that not all graphene-based materials exhibit FM-like features at room temperature. It depends upon the preparation process and edge atoms. In other words, one may infer that hydrogen termination facilitates ferromagnetism [1] where as the oxygen termination quenches magnetism [43].

Despite the clear phenomenological conclusions derived from the data in Figs.1 - 6, several important open questions remain. One unresolved issue is the chemical identity of the clusters involving carbon defects coupled with hydrogen. A related question is whether these clusters located at the edges of the ribbons or at the centers of the ribbons which cause ferromagnetism. A third outstanding question pertains to the relationship among $T_C$, cluster size, distance between two adatoms and defect concentration that would be anticipated from many models [32,23]. Experiments to address these important questions are underway.

To conclude, we have investigated the magnetism of GNRs and CCGNRs using the combination of magnetization and electron spin resonance measurements. Our classical MH measurements show the occurrence of ferromagnetlike features at low temperature for GNRs as well as for CCGNRs. However, such room temperature feature persists only for GNRs. While probing for in-depth understanding on the observed magnetism, our field cooled MH data infer the presence of anti-ferromagnetic regions mixed with ferromagnetic phase as evidenced from prominent exchange bias detected in both types of ribbons. Anti-ferromagnetic exchange coupling is found to exist between the FM and AFM phases in both types of ribbons, though GNRs show NEB whereas CCGNRs show PEB. Our electron spin resonance data infer that the carbon related localized states may be responsible for the observed magnetism. Furthermore, hyperfine sublevel correlation experiments performed on the former ribbons infer the presence of proton ($H^1$), may play a role in the observed magnetism, though the exact mechanism is far from clearly understood. ESR provides no indications for the presence of potassium (cluster) related signals, emphasizing the intrinsic magnetic nature of the ribbons. From our combined experimental findings, it may be inferred that the coexistence of ferromagnetic clusters with anti-ferromagnetic regions leading to disorder magnetism in both types of ribbons. Taking into account the numerous advantages of an organic carbon-based ferromagnetic-like magnetism together with semiconducting properties, such materials could be used for spin-based and magnet-electronic applications. Further experimental and theoretical efforts will not only bring us new insights into such novel materials but also pave the way to the new generation of molecule-based magnets.

**Acknowledgements**

SSR and SNJ would like to thank KU Leuven, for research fellowship. This work is supported by the K.U. Leuven Excellence financing (INPAC), by the Flemish Methusalem financing and by the IAP network of the Belgian Government. The National High Magnetic Field Laboratory is supported by NSF Cooperative Agreement No. DMR-0654118, and by the State of Florida.

## References:

1. D. Soriano, F. Muñoz-Rojas, J. Fernández-Rossier and J. J. Palacios, Phys. Rev. B 81, 165409 (2010).

2. O.V. Yazyev, Phys. Rev. Lett. 101, 037203 (2008).

3. L. R. Radovic and B. Bockrath, J. Am. Chem. Soc. 127, 5917 (2005).

4. F. J. Culchac, A. Latgé and A. T. Costa, New Journal of Physics 13, 033028 (2011).

5. H. Feldner, Z. Y. Meng, T. C. Lang, F. F. Assaad, S. Wessel, and A. Honecker, Phys. Rev. Lett. 106, 226401 (2008).

6. Y.W Son, M. L. Cohen and S. G. Louie, Nature 444, 347 (2006)

7. K. Sawada, F. Ishii, M. Saito, S. Okada, and T. Kawai, Nano Lett. 9, 269 (2009).

8. K. Wakabayashi, M. Fujita, H. Ajiki, and M. Sigrist, Phys. Rev.B 59, 8271 (1999); K. Kusakabe and M. Maruyama, Phys. Rev. B 67, 092406 (2003).

9. J. Červenka, M. I. Katsnelson and C. F. J. Flipse, Nature Physics 5, 840 (2009).

10. P. Esquinazi, R. H€ohne, K.-H. Han, A. Setzer, D. Spemann, T. Butz, Carbon 42, 1213 (2004).

11. Y. Wang, Y. Huang, Y. Song, X. Zhang, Y. Ma, J. Liang, Y.Chen, Nano Lett. 9, 220 (2009).

12. L. Xie, X. Wang, J. Lu, Z. Ni, Z. Luo, H. Mao, R. Wang, Y. Wang, H. Huang, D. Qi, R. Liu, T. Yu, Z. Shen, T.Wu, H. Peng, B. Özyilmaz, K. Loh, A. T. S. Wee, Ariando and W. Chen, Appl. Phys. Lett., 98, 193113 (2011); A. L. Friedman, H. Chun, Y. J. Jung, D. Heiman, E. R. Glaser, and L. Menon, Phys.Rev. B 81, 115461 (2010)

13. C. Tao, L. Jiao, O. V. Yazyev, Y-C Chen, J. Feng, X. Zhang, R. B. Capaz, J. M. Tour, A. Zettl, S. G. Louie, H.Dai and M. F. Crommie, Nature Physics 7,616 (2011).

14. H. Ohldag, T. Tyliszczak, R. Ho¨hne, D. Spemann, P. Esquinazi, M. Ungureanu, and T. Butz, Phys. Rev. Lett. 98, 187204 (2007); H Ohldag, P Esquinazi, E Arenholz, D Spemann, M Rothermel, A Setzer and T Butz, New Journal of Physics 12, 123012 (2010); J. Berashevich and T. Chakraborty, Nanotechnology 21, 355201 (2010).

15. N. Leconte, D. Soriano, S. Roche, P. Ordejon, J. C. Charlier and J. J. Palacios, ACS Nano. 5, 3987 (2011).

16. V. L. J. Joly, M. Kiguchi, S. J. Hao, K. Takai, T. Enoki, R. Sumii, K. Amemiya, H. Muramatsu, T. Hayashi, Y. A. Kim, M. Endo, J. C. Delgado, F. L. Urías, A. B. Méndez, H. Terrones, M. Terrones and M. S. Dresselhaus, Phys. Rev. B 81, 245428 (2010)

17. S. S. Rao, A. Stesmans, K. Keunen, D. V. Kosynkin, A. Higginbotham, J. M. Tour, Appl. Phys. Lett. 98, 083116 (2011).

18. D. V. Kosynkin, W. Lu, A. Sinitskii, G. Pera, Z. Sun, and J. M. Tour, ACS Nano. 5, 968 (2011).

19. D. V. Kosynkin, A. L. Higginbotham, A. Sinitskii, J. R. Lomeda, A. Dimiev, B. K. Price, and J. M. Tour, Nature (London) 458, 872 (2009); T. Shimizu, J. Haruyama, D. C. Marcano, D. V. Kosinkin, J. M. Tour, K. Hirose and K. Suenaga, Nature Nanotech. 249, 1 (2010); A. Sinitskii, A. A. Fursina, D. V. Kosynkin, A. L. Higginbotham, D. Natelson and J. M.Tour, Appl. Phys. Lett. 95, 253108 (2009).

20. A. Stesmans, Phys. Rev. B 48, 2418 (1993).

21. D.W. Boukhvalov and M. I. Katsnelson, ACS Nano. 4, 2440 (2011), and references therein.

22. A. V. Rode, E. G. Gamaly, A. G. Christy, J. G. F Gerald, S. T. Hyde, R. G. Elliman, B. Luther-Davies, A. I. Veinger, J. Androulakis, and J. Giapintzakis, Phys. Rev. B 70, 054407 (2004).

23. I. C. Gerber, A. V. Krasheninnikov, A. S Foster and R. M. Nieminen, New Journal of Physics 12, 113021 (2010).

24. Y. G. Semenov, J. M. Zavada, and K. W. Kim, Phys. Rev. B 84, 165435 (2011); M. Sepioni, R. R. Nair, S. Rablen, J. Narayanan, F. Tuna, R. Winpenny, A. K. Geim and I. V. Grigorieva, Phys. Rev. Lett., 105, 207205 (2010).

25. J Kunstmann, C. O¨ zdog˜an, A. Quandt and H. Fehske, Phys. Rev. B 83, 045414 (2011).

26. A. W. Mombrú, H. Pardo, R. Faccio, O. F. de Lima, E. R. Leite, G. Zanelatto, A. J. C.

Lanfredi, C. A. Cardoso and F. M. Araújo-Moreira, Phys. Rev. B 71, 100404 (R) (2005), and the references therein.

27. D. Arčon, Z. Jagličič, A. Zorko, A. V. Rode, A. G. Christy, N. R. Madsen, E. G. Gamaly, and B. Luther-Davies, Phys. Rev. B 74, 014438 (2006).

28. S. K. Mishra, F. Radu, H. A. Dürr, and W. Eberhardt, Phys. Rev. Lett., 102, 177208 (2009)

29. J. Nogues, D. Lederman, T. J. Moran and I. K. Schuller, Phys. Rev. Lett., 76, 4624 (1996)

30. X. Ke, M. S. Rzchowski, L. J. Belenky and C. B. Eom, Appl. Phys. Lett., 84, 5458 (2004); J. Appl. Phys. 97, 10K115 (2005).

31. D. Niebieskikwiat and M. B. Salamon, Phys.Rev.B 72, 174422 (2005)

32. L. D.Bianco, D. Fiorani, A. M. Testa, E. Bonetti and L. Signorini, Phys. Rev. B 70, 052401 (2004).

33. L. Pisani, B. Montanari and N. M. Harrison, New Journal of Physics 10, 033002 (2008).

34. R.C.Barklie, Diamond and Related Materials 10,174 (2001).

35. P. Hofer, A. Grupp, H. Nebenfuhr, M. Mehring, Chem. Phys. Lett. 132, 279 (1986).

36. F. Gerson and W. Huber, Electron Spin Resonance Spectroscopy of Organic Radicals; Wiley-VCH: New York, 2003.

37. C. I. Smith, H. Miyaoka, T. Ichikawa, M. O. Jones, J. Harmer, W. Ishida, P. P. Edwards, Y. Kojima and H. Fuji, J. Phys. Chem. C 113, 5409 (2009).

38. H.M. McConnell. J. Chem. Phys. 24, 632 (1956).

39. O.V. Yazyev, and M.I. Katsnelson, Phys. Rev. Lett., 100, 047209 (2008).

40. M. Fujita, K. Wakabayashi, K. Nakada, and K. Kusukabe, J. Phys. Soc. Jap 65, 1920 (1996).

41. Y.W Son, M. L. Cohen and S. G. Louie, Phys. Rev. Lett.,97, 216803 (2006).

42. O.V. Yazyev. Nano Lett 8, 1011 (2011).

43. R. G. A. Veiga, R. H. Miwa, and G. P. Srivastava, J. Chem. Phys. 128, 201101 (2008).

**Figure Captions:**

Fig. 1: (Color online) Isothermal magnetization (M-H) observed on GNRs collected from H of -5 to + 5T at various temperatures 5, 10, 50, 100, 200 and 300 K (a). The inset of this figure presents the enlarged version of M-H curve. The lower right inset of this figure represents the Langevin fit (solid curve) to the experimental data using equation (1). Isothermal magnetization (M-H) observed on CCGNRs collected from H of -5 to + 5 T at various temperatures 1.8, 5, 10, and 15 K (b). The top left inset of this figure shows the enlarged version of M-H curve. The lower right inset of this figure represents the Langevin fit (solid curve) to the experimental data using equation (1), see text for more details. In the former ribbons, a clear ferromagnetic character can be seen even at room temperature (300 K). However, in the later case, a clear M-H open loop is visible at 1.8 K.

Fig. 2: (Color online) A temperature dependence of zero-field cooled and field-cooled dc magnetic susceptibilities for (a) GNRs (b) CCGNRs. In both cases, for field cooled measurements, the applied magnetic field was 0.1 T.

Fig. 3: (Color online) Isothermal M-H curves collected from GNRs (a) and CCGNRs (b) under zero field and field cooled conditions (0.1, 0.3, 0.5 and 1 T; 0.1, 0.2, 0.4, 0.6 T), respectively. The top left insets are the corresponding enlarged versions inferring the negative exchange bias in the former case (a), and the positive exchange bias in the later case (b). The solid curves in the bottom right insets are resulted from the optimized computer fittings using equation (2), see text for more details.

Fig. 4: (Color online) Variation of coercive field ($H_c$) as a function of cooling magnetic field ($H_{FC}$) measured at 5 K for (a) GNRs (b) CCGNRs. The curves are the guide to the eye.

Fig. 5: First derivative CW K-band ESR spectra measured at 4.2 K on (a) GNRs (b) CCGNRs, using $B$m= 0.42 × 10$^{-4}$ T and $P$μ = 2.5 $nW$. The signal at $g \approx 1.99869$ stems from a co-mounted Si:P marker sample.

Fig. 6: (Color online) An X-band 2D plot of HYSCORE spectrum collected at 12 K on GNRs
in frequency coordinates

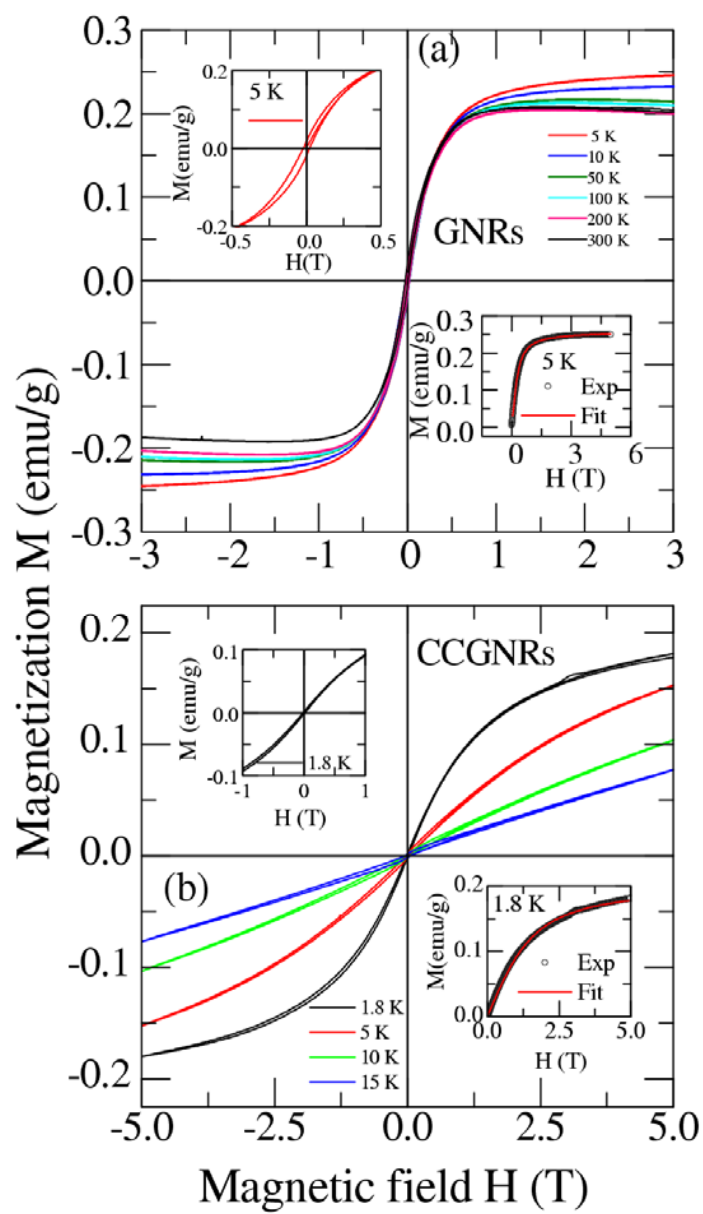

Fig. 1: (Color online) Isothermal magnetization (M-H) observed on GNRs collected from H of -5 to + 5T at various temperatures 5, 10, 50, 100, 200 and 300 K (a). The inset of this figure shows the enlarged version of M-H curve. The lower right inset of this figure represents the Langevin fit (solid curve) to the experimental data using equation (1). Isothermal magnetization (M-H) observed on CCGNRs collected from H of -5 to + 5 T at various temperatures 1.8, 5, 10, and 15 K (b). The top left inset of this figure shows the enlarged version of M-H curve. The lower right inset of this figure represents the Langevin fit (solid curve) to the experimental data using equation (1), see text for more details. In the former ribbons, a clear ferromagnetic character can be seen even at room temperature (300 K). However, in the later case, a clear M-H open loop is visible at 1.8 K.

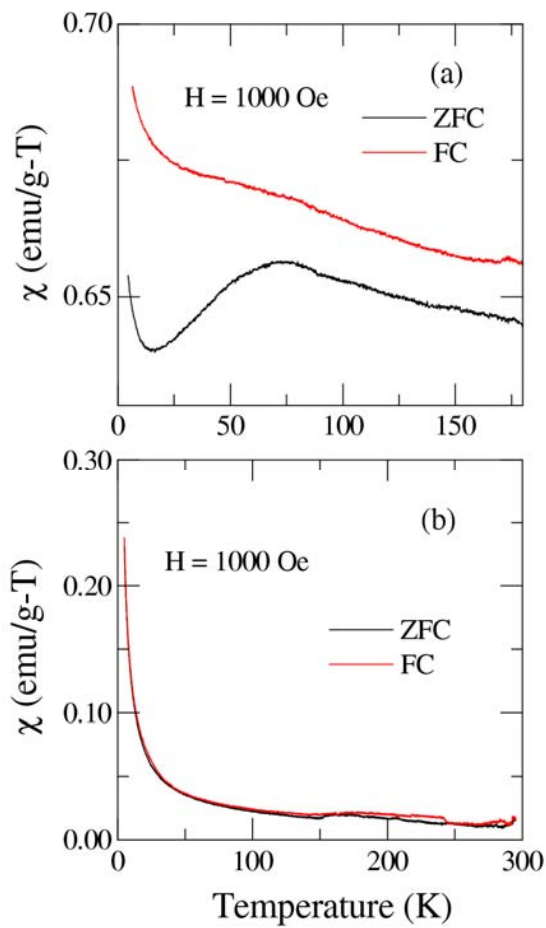

Fig. 2: (Color online) A temperature dependence of zero-field cooled and field-cooled dc magnetic susceptibilities for (a) GNRs (b) CCGNRs. In both cases, for field cooled measurements, the applied magnetic field was 0.1 T.

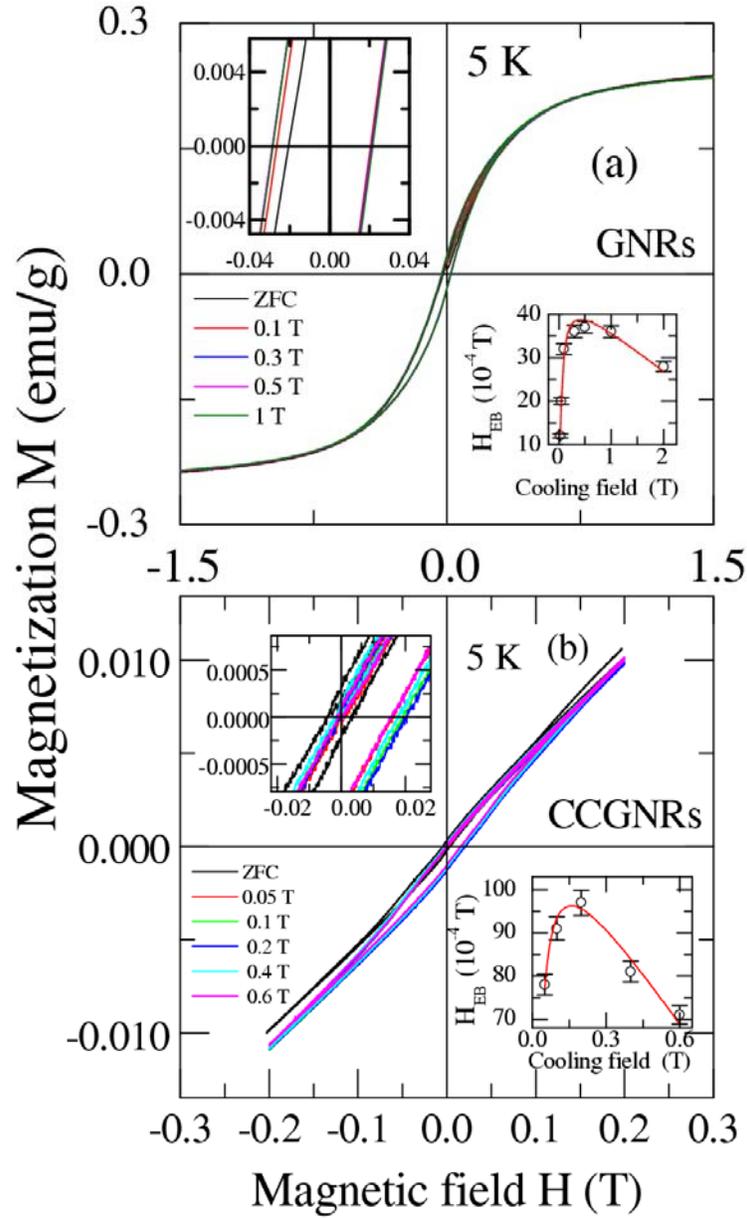

Fig. 3: (Color online) Isothermal M-H curves collected from GNRs (a) and CCGNRs (b) under zero field and field cooled conditions (0.1, 0.3, 0.5 and 1 T; 0.1, 0.2, 0.4, 0.6 T), respectively. The top left insets are the corresponding enlarged versions inferring the negative exchange bias in the former case (a), and the positive exchange bias in the later case (b). The solid curves in the bottom right insets are resulted from the optimized computer fittings using equation (2), see text for more details.

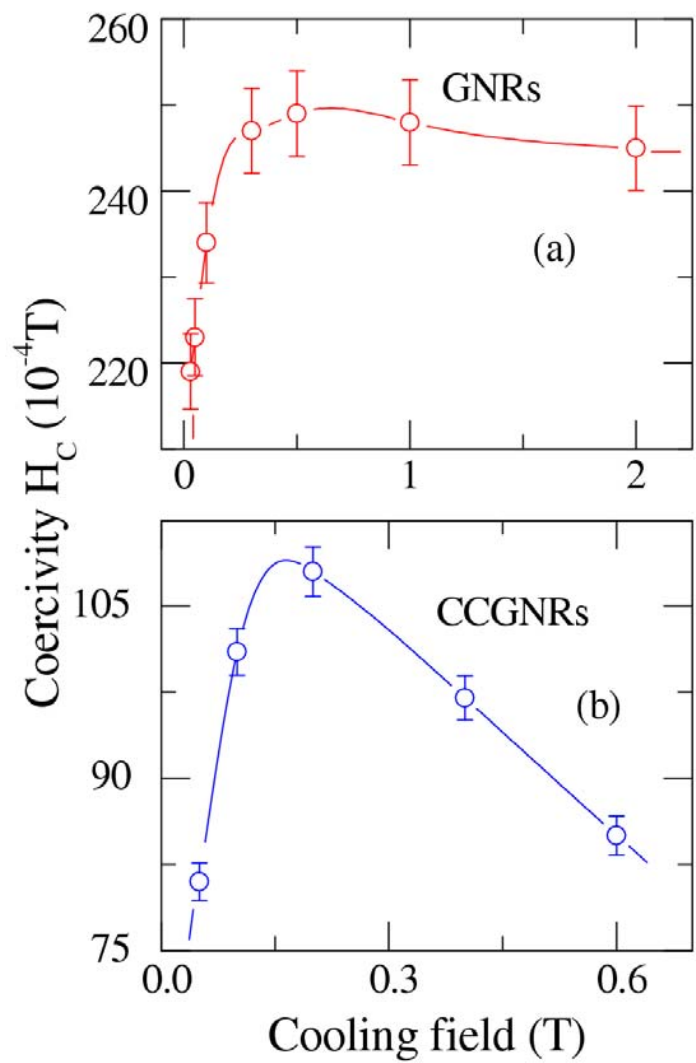

Fig. 4: (Color online) Variation of coercive field ($H_c$) as a function of cooling magnetic field ($H_{FC}$) measured at 5 K for (a) GNRs (b) CCGNRs. The curves are guide to the eye.

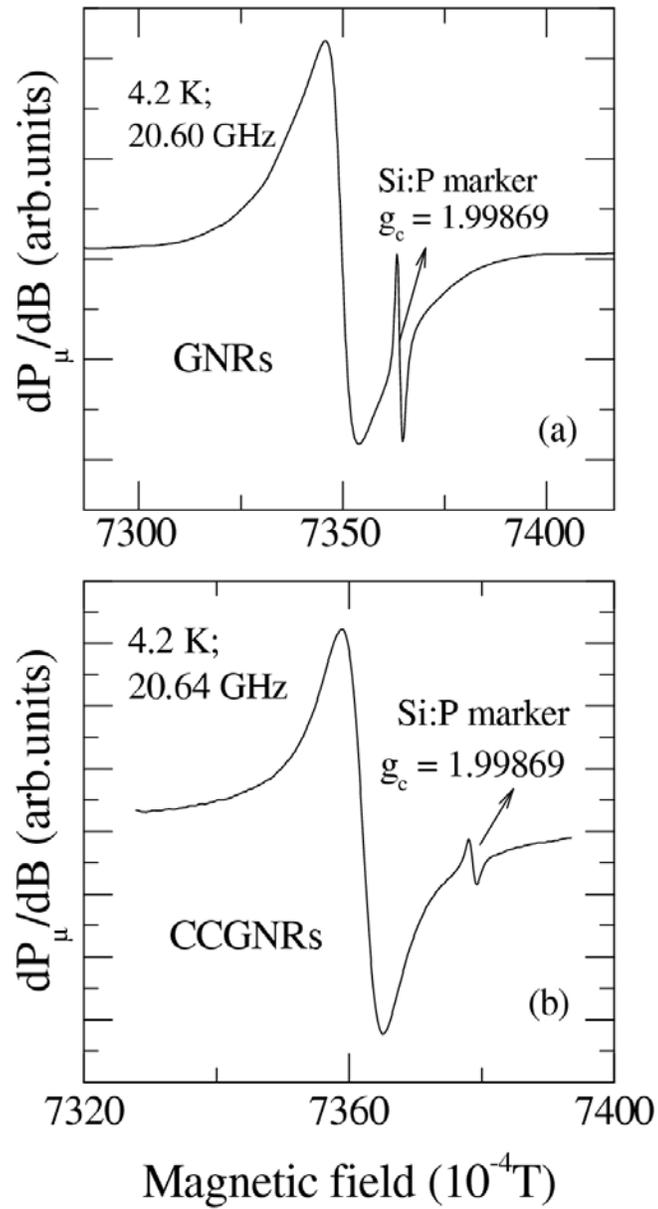

Fig. 5: First derivative CW K-band ESR spectra measured at 4.2 K on (a) GNRs (b) CCGNRs, using $Bm = 0.42 \times 10^{-4}$ T and $P\mu = 2.5$ *nW*. The signal at g $\approx 1.99869$ stems from a co-mounted Si:P marker sample.

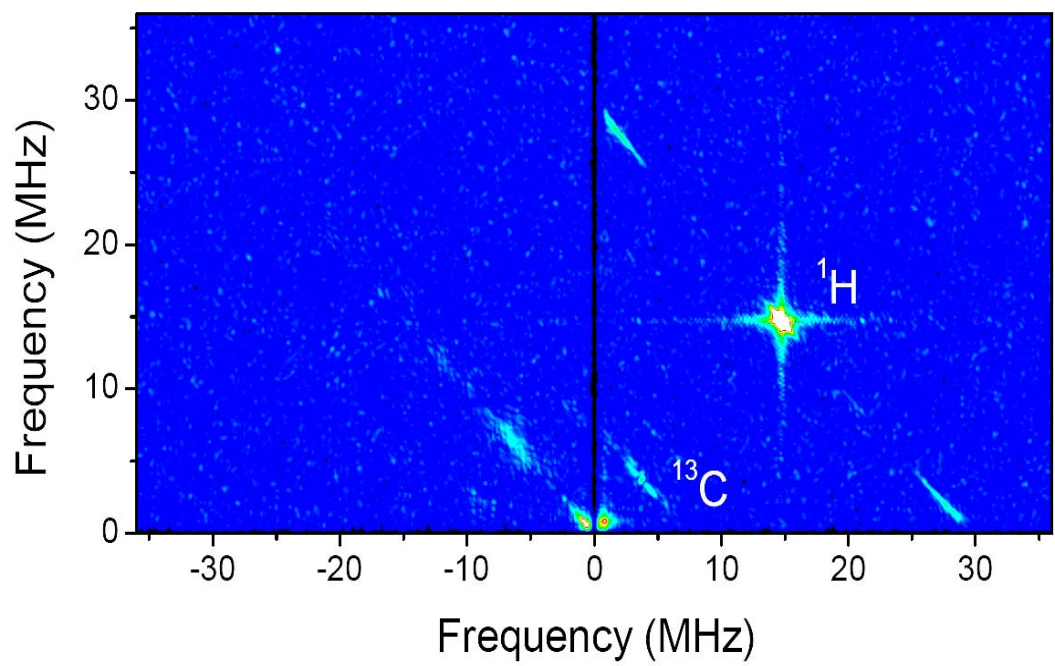

Fig. 6: (Color online) An X-band 2D plot of HYSCORE spectrum collected at 12 K on GNRs in frequency coordinates